\documentclass[12pt,preprint]{aastex}

\shorttitle{Geodetic Precession in PSR J1141--6545}
\shortauthors{Hotan et al.}

\begin{document}


\title{Geodetic Precession in PSR J1141--6545}


\author{A. W. Hotan}
\affil{Swinburne University of Technology}
\email{ahotan@astro.swin.edu.au}

\author{M. Bailes}
\affil{Swinburne University of Technology}
\email{mbailes@swin.edu.au}

\author{S. M. Ord}
\affil{Swinburne University of Technology}
\email{sord@swin.edu.au}


\begin{abstract}
We present observations that show dramatic evolution of the mean pulse
profile of the relativistic binary pulsar J1141--6545 over a period of
five years. This is consistent with precession of the pulsar spin axis
due to relativistic spin-orbit coupling. Observations made between
1999 and 2004 with a number of instruments at the Parkes radio
telescope demonstrate a steady, secular evolution of the mean total
intensity profile, which increases in width by more than 50 percent
during the five year period. Analysis of the changing position angle
of the linearly polarised component of the mean profile suggests that
our line of sight is shifting closer to the core of the emission
cone. We find that the slope of the position angle swing across the
centre of the pulse steepens with time and use a simplified version of
the rotating vector model to constrain the magnitude and direction of
the change in our line of sight angle relative to the pulsar magnetic
axis. The fact that we appear to be moving deeper into the emission
cone is consistent with the non-detection of this pulsar in previous
surveys.
\end{abstract}



\section{Introduction}
\label{sec:intro}

Soon after the discovery of the binary pulsar B1913+16 \citep{ht74} it
was pointed out that if the pulsar's spin axis and orbital angular
momentum vector were misaligned, they should precess around their
common sum on a time scale of around 300 years due to General
relativistic spin-orbit coupling \citep{dr74,hdr75,bo75,eh75}. This
phenomenon is often referred to as ``geodetic precession''.  Because
radio pulsars are thought to possess lighthouse-like beams of emission
that only beam to a small fraction of the sky, such precession should
lead to observable pulse shape changes once a significant fraction of
the precession period has passed.

Secular changes in the mean pulse profile of PSR B1913+16 were first
reported by \citet{wrt89}. The changes became more pronounced in later
years, allowing limited modeling of the emission cone
\citep{kra98,wt02}. The variations seen in PSR B1913+16 suggest a
small misalignment angle between the spin and orbital angular momentum
vectors, consistent with a natal kick being imparted to the most
recently formed neutron star \citep{bai88}. Large misalignment angles
raise the possibility that we might be able to, for the first time,
map out the entire emission cone of a pulsar until it disappears
completely from view \citep{kra98}.

For many years, PSR B1913+16 was the only binary pulsar with the right
combination of relativistic parameters and an evolutionary history
that would make detection of geodetic precession possible. However,
shortly after geodetic precession was first hinted at, new
relativistic pulsars were uncovered in pulsar surveys that would
become suitable for measurements in the future. \citet{agk+90}
discovered a near clone of the original binary pulsar in the globular
cluster M15C, which although weak, might one day exhibit the
phenomenon. The much closer PSR B1534+12 \citep{wol90a} is a
relativistic pulsar in a 10 hour binary orbit that has recently been
shown to exhibit pulse shape changes. These changes have been combined
with polarimetric models to make the first reliable estimate of the
rate at which the spin axis of the pulsar is tilting away from us
\citep{sta04}. Thus, measurements of geodetic precession are allowing
new tests of General relativity and constraining evolutionary models
\citep{wkk00,kws03}. These new tests complement the earlier pioneering
work of authors such as \cite{tw82}, that examined other aspects of
the theory.

In principle, it should be possible to use the relativistic pulsars to
map both shape and intensity changes across a pulsar emission
cone. However, this is complicated by the fact that many of the
pulsars have random variations in their total intensity because of
refractive and diffractive interstellar scintillation. Pulsars with
small dispersion measures (DM) often have their fluxes change by
factors of several on time scales that vary between minutes and days
\citep{ssh+00}. However, large dispersion measure pulsars ($DM > 100$
pc cm$^{-3}$) have relatively stable fluxes when observed at high
frequencies ($\nu > $ 1.4 GHz) if long integration times are used and
the observing system samples a large ($> 100$ MHz) bandwidth.  The
discovery of a relativistic pulsar at a large DM with a rapid geodetic
precession time scale might then offer the first hope of determining
how rapidly pulsar emission varies in both intensity and shape across
the emission cone. This discovery would have important ramifications
for pulsar statistics, where one of the great uncertainties is the
pulsar ``beaming fraction'', often guessed at by observing the duty
cycles of radio pulsars. Various arguments have been made in favour of
both meridional compression \citep{big90b} and the elongation of
pulsar beams \citep{nv83}, making the actual measurement very
important.

In the future it seems as though we will have a much larger number of
relativistic systems with which to both test General relativity and
map pulsar emission cones. Recent surveys have been finding
relativistic pulsars at a welcome rate \citep{fsk+04} and the total
pulsar population now exceeds 1500 objects. In particular, the Parkes
multibeam surveys have discovered a large number of pulsars in
relativistic orbits. The spectacular ``double pulsar'' is a 2.4 hour
binary with two active pulsars, and a geodetic precession period of
just 80 years \citep{lbk+04}. \cite{fkl+05} report the recent
discovery of the 7.7-hour binary pulsar J1756--2251 with an
eccentricity of 0.18 that almost certainly consists of two neutron
stars.

The first relativistic binary pulsar discovered by the multibeam
surveys was however the 4.8 hour binary pulsar J1141--6545
\citep{klm+00}. This pulsar orbits what is most likely a white dwarf
companion, but the system is unique in that it still possesses a
significant orbital eccentricity ($e=0.17$). An eccentricity of this
magnitude suggests that the system was put in its final configuration
in an explosive event that may have given the pulsar a significant
kick. Recent timing of the system is consistent with a 1.3 M$_\odot$
pulsar orbiting a 1.0 M$_\odot$ white dwarf companion
\citep{bok03}. It is likely that the system originated as a binary
containing two main sequence stars, both below the critical mass
required for a supernova. The initially more massive star begins to
transfer matter onto its companion as it evolves, causing it to exceed
the critical mass. If the system remains bound after the resulting
supernova, a young neutron star is left orbiting a white dwarf
companion. In the case of symmetric supernovae, the eccentricity of
the orbit is induced by the sudden mass loss, and allows us to
determine the pre-supernova mass uniquely \citep{rs85}. This ejected
mass can be related to the expected runaway velocity of the system,
which in the case of PSR J1141--6545, is less than 50 km s$^{-1}$.

Significant progress has been made in understanding the geometry and
location of PSR J1141--6545 through a range of timing and
spectroscopic studies. \citet{obv02a} demonstrated via HI absorption
analysis that the pulsar is at least 3.7 kpc distant. In addition, PSR
J1141--6545 was the first pulsar to exhibit dramatic changes in its
scintillation time scale as a function of orbital phase, which have
enabled an independent estimate of both its orientation and
velocity. \citet{obv02} used the orbital modulation of the
scintillation time scale to calibrate the usually unknown scale factor
that relates the scintillation time scale, scattering screen distance
and intrinsic motion to pulsar velocity. Their subsequently determined
space motion was greater than that expected from a symmetric supernova
\citep{obv02}.

Several years ago we commenced an observing campaign of PSR
J1141--6545 to study its HI, scintillation and timing properties over
long baselines. It was clear that such observations could also be used
to search for relativistic effects such as orbital decay and
precession. The DM of the pulsar is 116 pc cm$^{-3}$, and its
scintillation properties are well understood. \citet{obv02} showed
that the scintillation bandwidth is much smaller than the 256 MHz
observing band used by the analogue filter bank at the Parkes radio
telescope, resulting in fairly stable fluxes when integrated over the
4.8 hr orbit.

In this paper we demonstrate that PSR J1141--6545 is undergoing rapid
secular evolution of both its total intensity and polarimetric
profiles in a manner consistent with geodetic precession. In section
\ref{sec:observations} we describe the many instrumental systems used
to observe this pulsar since its discovery in 1999, along with the
associated data reduction methods. Section \ref{sec:1141} describes
the parameters of the binary system in greater detail and includes a
calculation of the expected precession rate. It also introduces two
polarimetric profiles that are considerably different from each other,
providing the first evidence of profile evolution. Section
\ref{sec:tip_change} describes in detail the observed secular changes
in the total intensity profile. Polarimetric evolution is considered
in section \ref{sec:pol_change}, which shows that there has been a
convergence of the linear and circular components of the pulsar
profile in the last twelve months and that the slope of the position
angle swing is steepening, presumably as we approach the emission pole
of the pulsar. Finally, in section \ref{sec:discussion} we discuss the
implication of our observations for pulsar surveys and pulsar emission
models.

\section{Observations}
\label{sec:observations}

Observations were made at the Parkes radio telescope between July 1999
and May 2004, at centre frequencies ranging from 1318.25 MHz to
1413.50 MHz. Two different receiver packages were used to record data
during this period; the central beam of the Parkes multibeam receiver
and the wide band H-OH receiver. The multibeam receiver \citep{swb+96}
has a system temperature of approximately 21K at 20 cm, which was
about 5 degrees cooler than the H-OH receiver before it was upgraded
near the end of 2003. Flux calibration of both systems using the radio
galaxy 3C218 (Hydra A) suggests that the post-upgrade difference is
only one or two degrees Kelvin. Our data were recorded with three
different instruments, each designed for high time resolution
observations across the widest possible bandwidths. Due to the rapid
development of digital electronics within the past decade, each new
observing system differed significantly from its predecessor.

The Caltech Fast Pulsar Timing Machine (FPTM), described by
\citet{nav94}, was a hardware-based auto-correlation spectrometer that
performed incoherent dedispersion of dual orthogonal polarisations
across two bands, each 128 MHz wide. Although the sampling rate was
high enough to observe millisecond pulsars (MSPs) with only a few
microseconds of smearing (at low DM), this instrument suffered from a
number of artifacts induced by radio frequency interference and some
deterioration in the correlator boards themselves. In some pulsars
this led to oscillations in the passband that contaminated the pulsar
profile. Nevertheless, many observations with this instrument were not
affected by these problems and it successfully timed many MSPs to high
accuracy \citep{tsb+99}. The FPTM 2-bit sampled the raw data and
formed auto-correlation functions that were binned at the apparent
spin period of the pulsar. We were able to apply routine 2-bit
corrections to enable accurate polarimetry. Being an incoherent
detector with a finite number of lags, the FPTM could divide the
passband into 4$\times$128$\times$1 MHz channels, leaving a residual
dispersion smear ($t_{\rm{smear}}$) given by Eq \ref{eq:smear}.

\begin{equation}
t_{\rm{smear}} = {8.3}\frac{B DM}{\nu^3}  \mu s
\label{eq:smear}
\end{equation}

\noindent Here, the channel bandwidth $B$ is in units of MHz, the sky
frequency $\nu$ is in units of GHz and the dispersion measure $DM$
is in units of pc cm$^{-3}$. For the configuration used to observe PSR
J1141--6545, this corresponds to 350 $\mu$s of smearing in the
detected pulse profile. Given that the FPTM uses 1024 phase bins
across a single pulse period and that PSR J1141--6545 rotates once
every 394 ms, each phase bin represents 384 $\mu$s of time. The
detrimental effects of dispersion smearing are therefore confined to
within a single phase bin.

However, if the pulsar spin period is two orders of magnitude shorter
(as is typical of the millisecond pulsar population), dispersion
smearing can significantly reduce the resolution of an incoherent
detector. Motivated by a desire to overcome this problem for MSPs, the
Caltech Parkes Swinburne Recorder Mk I (CPSR1) was commissioned in
1998 August. This system implemented a technique called coherent
dedispersion \citep{hr75}, which requires Nyquist sampling of the
observed band, followed by deconvolution with a response function
characteristic of the interstellar medium (ISM). While this approach
effectively eliminates dispersion smearing in the detected profiles,
it is highly computationally intensive both in terms of the initial
data rate and subsequent analysis. CPSR1 streamed digital samples to a
striped set of four DLT drives (analogous to the method implemented
for the S2 VLBI recorder) whose tapes were shipped to the Swinburne
Centre for Astrophysics and Supercomputing for processing. Even with
four striped tape drives, CPSR1 was limited to a bandwidth of 20
MHz. Rapid advances in consumer digital electronics soon made it
feasible to upgrade the capabilities of the system and in 2002 August,
CPSR Mk II was installed at Parkes. CPSR2 performs coherent
dedispersion in near real-time, using a cluster of 30 server-class
computers located at the telescope. It is capable of recording
2$\times$64 MHz dual-polarisation bands simultaneously, providing a
total bandwidth comparable to that of the previous generation of
incoherent detectors, like the FPTM. The coherent dedispersion method
employed by both CPSR machines allows essentially arbitrary spectral
resolution and reduces the dispersion smearing in each channel to a
minute fraction of PSR J1141--6545's period, giving an effective
sampling time of a few microseconds.

Individual observations of PSR J1141--6545 ranged in duration from a
few minutes to several hours. In recent years, our strategy has been
to maximise orbital phase coverage by observing in concentrated
sessions during which the pulsar is tracked continuously for two whole
orbits ($\sim$9.6 hr).  To calibrate the data we point the telescope
one degree south of the pulsar and drive the in-built receiver noise
source with a square wave at a frequency of 11.122976 Hz at least once
per orbit, to characterise the polarimetric response of the system. In
addition, at least once per month we observe the flux calibration
source 3C218 (Hydra A).

Our highest density of observations were taken with CPSR2 during 2003
and 2004 (MJD 52845 - 53134), during which time we have a full record
of polarimetry, flux and profile morphology. The CPSR1 recorder was
designed primarily to observe the bright southern millisecond pulsar
J0437--4715 \citep{vbb+01}, however in 2001 January it took data on PSR
J1141--6545 for a total of 30 hours beginning on MJD 51922. The
resulting calibrated profile provides important, early epoch
information. We have selected three profiles at the widely spaced
epochs of MJD 51381, 51781 and 52087, representing high quality FPTM
data to further supplement our temporal coverage. It is fortuitous
that our earliest FPTM pointing (MJD 51381) dates all the way back to
1999, extending our time baseline by almost two full years. For this
reason, we include a 1999 profile despite the fact that the
observation was only 12 minutes in duration. Fortunately, PSR
J1141--6545 is a bright pulsar with an average flux density of
approximately 4 mJy, so the signal-to-noise (S/N) ratio of this 12
minute observations is 125, quite sufficient for our analysis.

All data were processed using the tools included with the PSRCHIVE
\citep{hvm04} scheme, with the addition of several extra routines
specific to pulse variability analysis\footnote{All PSRCHIVE code is
freely available for academic use, see
http://astronomy.swin.edu.au/pulsar}.

\section{PSR J1141--6545}
\label{sec:1141}

PSR J1141--6545 was discovered in the first Parkes multibeam survey
\citep{klm+00a}. It resides in an unusual relativistic binary system,
orbiting what is most likely a heavy white dwarf companion once every
4.8 hours. The pulsar does not appear to be recycled and is assumed to
be the most recently evolved member of the system. Positioned close to
the galactic plane, it is one of the few pulsars whose distance can be
estimated by analysis of neutral hydrogen absorption features in its
frequency spectrum. \citet{obv02a} obtain a lower limit of 3.7 kpc
using this method. In addition, the signal from this pulsar exhibits
diffractive scintillation over small bandwidths ($\sim$1 MHz) and time
scales of a few minutes \citep{obv02}, which can be used to place
timing-independent constraints on the binary parameters. \citet{obv02}
report a significant detection of orbital modulation in the observed
scintillation velocity (due to the motion of the pulsar in its orbit)
and infer both a relative velocity of $\sim$115 km s$^{-1}$ and an
orbital inclination angle $i$ = 76 $\pm$ 2.5$^\circ$. This velocity is
large compared to the value of $<$ 50 km s$^{-1}$ expected to result
from a symmetric supernova. Uncertainties in our knowledge of the true
distance to the pulsar, and hence the relative velocity of the Earth's
standard of rest, combined with uncertainties introduced by the
Earth's orbital motion and bulk flows or anisotropies in the ISM, mean
that we cannot convincingly state that the pulsar has an anomalous
velocity due to an asymmetric supernova. However, if the profile
evolution reported here is due to spin and orbital angular momentum
misalignment and geodetic precession, we would expect the pulsar to
have received a kick at birth, which would increase its runaway
velocity over that of a symmetric explosion.

The rotation period of PSR J1141--6545 is 394 ms, so the precision
obtainable through pulse timing experiments is somewhat limited when
compared to results \citep{vbb+01} obtained by timing millisecond
pulsars, whose spin periods are of order 100 times shorter. Despite
this, several post-Keplerian parameters are measurable. \citet{bok03}
describe a timing solution that includes significant detections of
periastron advance ($\dot \omega \sim$ 5.3$^\circ$ yr$^{-1}$),
combined transverse Doppler and gravitational redshift ($\gamma$) and
a marginal detection of orbital period derivative ($\dot P_{\rm
b}$). Despite the lack of any Shapiro delay measurement, we can still
derive a good estimate of the component masses. The post-Keplerian
parameters $\dot \omega$ and $\gamma$ are related to the pulsar mass
($m_{\rm p}$) and companion mass ($m_{\rm c}$) by Eq \ref{eq:omdot}
and Eq \ref{eq:gamma} respectively. In addition, pulse timing
accurately determines the quantity, derived from Newtonian
gravitation, known as the mass function (Eq \ref{eq:massfn}). These
equations allow determination of the inclination angle and component
masses.

\begin{equation}
\dot \omega = 3 \left({{2 \pi}\over{P_{\rm b}}}\right)^{5/3}
\left({{G (m_{\rm p} + m_{\rm c})}\over{c^3}}\right)^{2/3}
(1-e^2)^{-1}
\label{eq:omdot}
\end{equation}

\begin{equation}
\gamma = e \left({\frac{P_{\rm b}}{2 \pi}}\right)^{1/3} 
\frac{G^{2/3}}{c^{2}} m_{\rm c} (m_{\rm p} + 2 m_{\rm c}) 
{(m_{\rm p} + m_{\rm c})}^{4/3}
\label{eq:gamma}
\end{equation}

\begin{equation}
f(m_{\rm p},m_{\rm c}) = {{m_{\rm c}^3} \sin^3
i\over{(m_{\rm p} + m_{\rm c})^2}} = {{4 \pi^2}\over{G}}{{a^3 \sin^3
i}\over{P_{\rm b}^2}}
\label{eq:massfn}
\end{equation} 

\noindent Here, $G$ is Newton's gravitational constant, $P_{\rm b}$ is
the pulsar orbital period and $a \sin i$ is the projected semi-major
axis.  According to \citet{bok03}, $m_{\rm p}$ = 1.30 $\pm$ 0.02
M$_\odot$ and $m_{\rm c}$ = 0.986 $\pm$ 0.02 M$_\odot$. The
timing-derived inclination angle limit ($i > 75^\circ$) compares well
with the value obtained from scintillation experiments \citep{obv02}.

Assuming that General relativity is the correct theory of gravity,
\citet{bo75b} present an expression (Eq \ref{eq:prate}) for the
expected, time-averaged precession rate of the pulsar spin axis,
$\Omega_{\rm p}$.

\begin{equation}
\Omega_{\rm p} = {\frac{1}{2}}
{\left(\frac{G}{c^3}\right)}^{2/3}
{\left(\frac{P_{\rm b}}{2\pi}\right)}^{-5/3}
{\frac{m_{\rm c}(4m_{\rm p}+3m_{\rm c})}{(1-e^2)(m_{\rm p}+m_{\rm c})^{4/3}}}
\label{eq:prate}
\end{equation}

\noindent Here, $c$ is the speed of light and $e$ is the eccentricity
of the system. For PSR J1141--6545, this evaluates to an intrinsic
precession rate of 1.36$^\circ$ yr$^{-1}$, which implies a precession
period of 265 years.

\citet{bai88} showed graphically that the maximum observable rate of
precession may be significantly less than the intrinsic value. The
geometry of the system and our viewing angle have a significant impact
on our ability to detect geodetic precession. The observable quantity
is the rate at which the angle $\delta$ between the observer's line of
sight and the pulsar spin axis changes, as this will manifest itself
as a changing cut through the emission cone. \citet{bai88} and
\citet{cwb90} present expressions for the rate of change of $\delta$,
of the form reproduced in Eq \ref{eq:delta}. The most important
parameters in the expression are the misalignment angle between the
spin axis and the orbital angular momentum vector, and the
precessional phase at the current epoch, neither of which are known.

\begin{equation}
d\delta/dt = 
\Omega_{p} {\bf n} \cdot ({\bf s} \times {\bf j})
(1 - [{\bf n} \cdot {\bf s}]^{2})^{-1/2}
\label{eq:delta}
\end{equation}

\begin{centering}
\begin{figure}
\includegraphics[angle=270,scale=0.65]{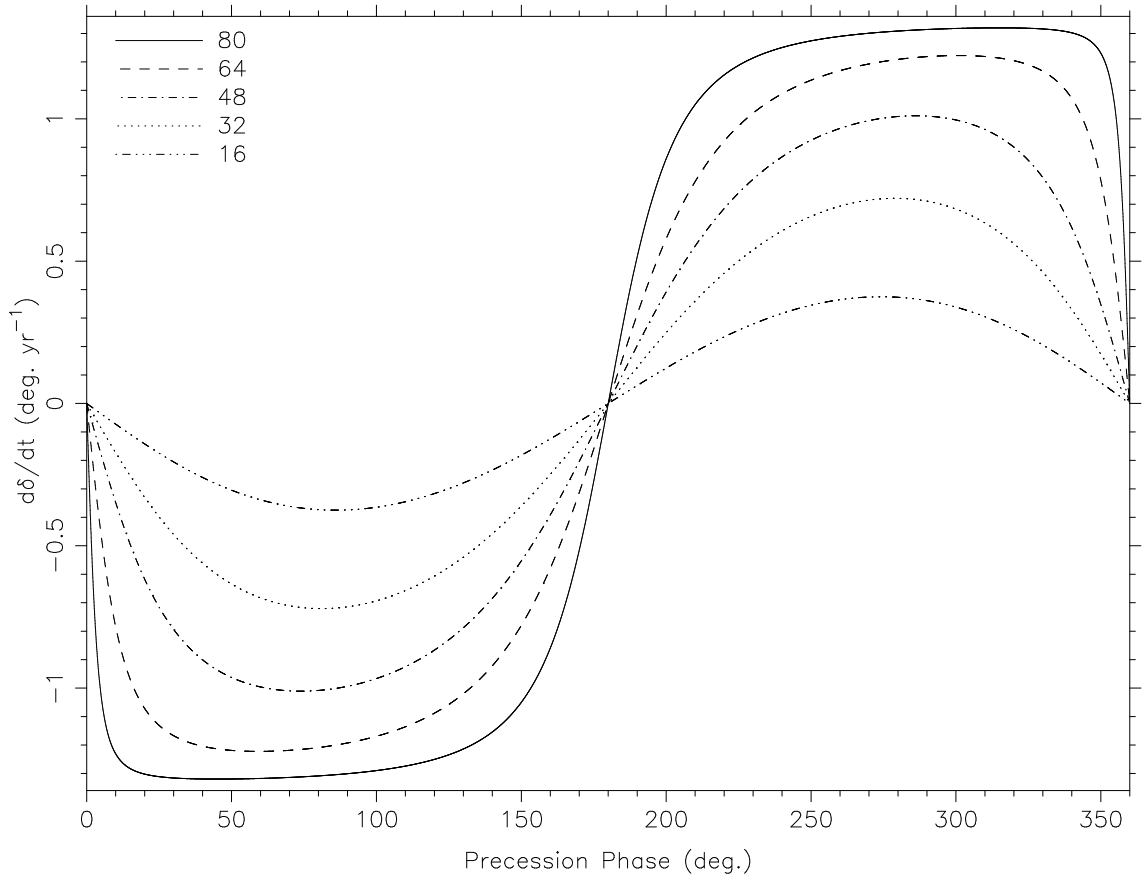}
\caption{Rate of change of the angle between our line of sight and the
spin axis of the pulsar ($d\delta/dt$) as a function of precessional
phase, for various misalignment angles (shown inset top left, in
degrees) for PSR J1141--6545. We have assumed an inclination angle of
76 degrees, consistent with scintillation measurements. The amplitude
of the observable precession signature is highly dependent on both the
misalignment angle and our current position in the precession cycle,
but any derived value of $d\delta/dt$ gives a minimum misalignment
angle for the system.}
\label{fig:delta}
\end{figure}
\end{centering}

\noindent Here, ${\bf n}$ is a unit vector along the line of sight to
the observer, ${\bf s}$ is a unit vector along the pulsar spin axis
and ${\bf j}$ is a unit vector in the direction of the orbital angular
momentum. To evaluate this equation, we must know the misalignment
angle, the current precessional phase and the orbital inclination
angle $i$. The first two parameters are unknown for the PSR
J1141--6545 system, but we can assume the value of $i$ derived from
scintillation studies and plot one precessional period of $d\delta/dt$
for various misalignment angles (see Fig \ref{fig:delta}).

Unfortunately, it is difficult to translate an observed difference in
mean pulse profile morphology into a quantitative measure of angular
shift. This requires knowledge of the intrinsic beam shape, which of
course we do not possess, making the result model dependent. It is
possible that polarimetric studies may offer a key alternative method,
however it is still necessary to assume some model of pulsar
polarisation as a function of impact angle. In this paper we restrict
ourselves, where possible, to a quantitative description of the
observed evolution of the mean pulse profile.

\subsection{Selected Observations}

\begin{centering}
\begin{figure}
\includegraphics[angle=270,scale=0.65]{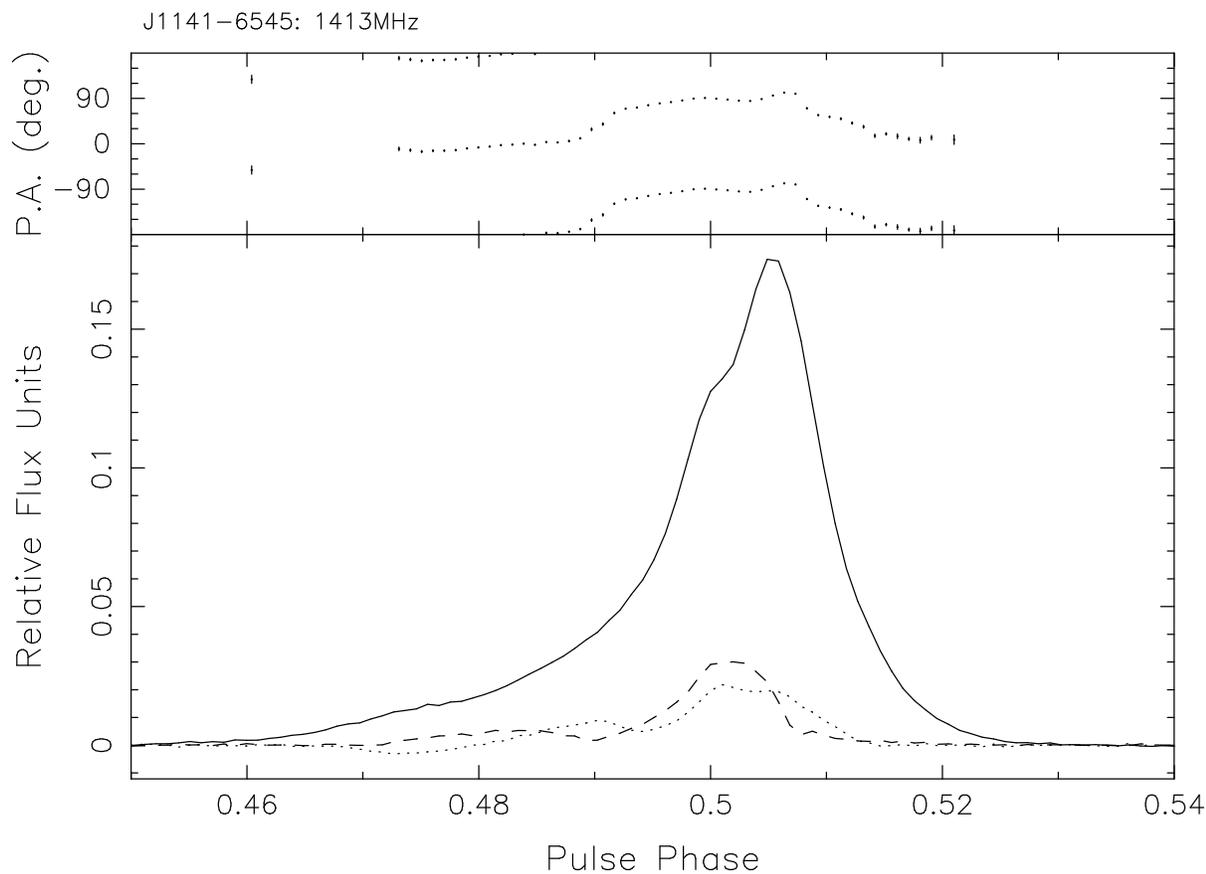}
\caption{PSR J1141--6545 mean profile, obtained from 28 hours of data
taken in 2001 January (MJD 51922), at a centre frequency of 1413
MHz. The solid line represents total intensity, the dashed line total
linear and the dotted line total circularly polarised emission. There
are 1024 phase bins across the profile, which has been
polarimetrically calibrated using a simple model of relative gain and
phase in the orthogonal linear receiver probes. Note the slight
``bump'' on the leading edge of the profile and the absence of any
steep position angle evolution across the phase range shown.}
\label{fig:cpsr1}
\end{figure}
\end{centering}

Fig \ref{fig:cpsr1} shows a coherently dedispersed, polarimetrically
calibrated mean profile, observed with the multibeam receiver and
CPSR1 at a frequency of 1413 MHz in 2001 January (MJD 51922). The
pulse profile is morphologically quite simple, consisting of a single
component flanked on the left by a shoulder of emission. Note the
small ``bump'' high on the leading edge of the profile, which is also
present in the FPTM and analogue filter bank data. The peak fractional
polarisation is of order 20 percent in both linear and circular. The
position angle swing does not seem to fit the predictions of the
rotating vector model and is similar to that seen by \citet{klm+00},
although lacking the orthogonal mode change that is present in the
leading shoulder of the Kaspi et al. profile.

\begin{centering}
\begin{figure}
\includegraphics[angle=270,scale=0.65]{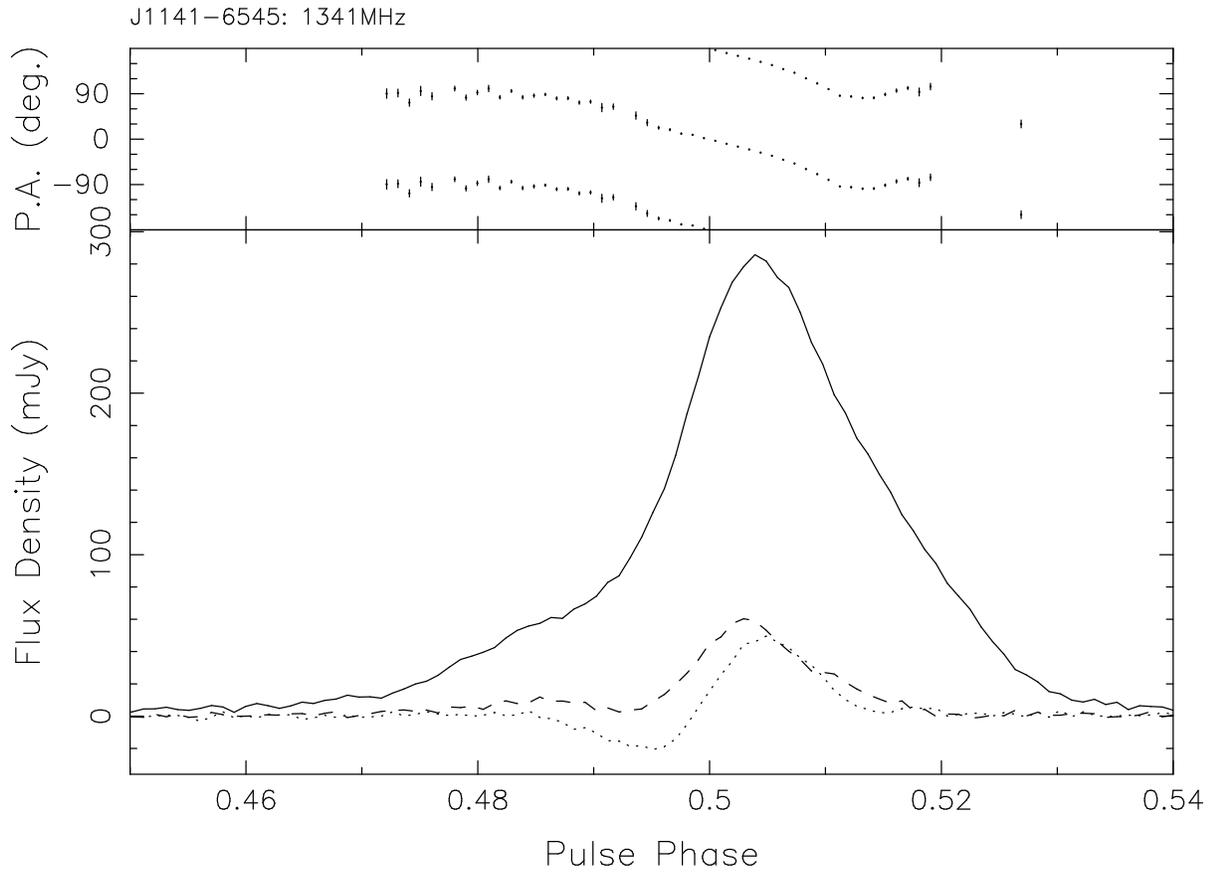}
\caption{2004 May (MJD 53134) PSR J1141--6545 profile (as in Fig
\ref{fig:cpsr1}) obtained from 2 hours of data taken with CPSR2 at a
centre frequency of 1341 MHz. Note the smooth leading edge and
extended trailing component, as well as the more pronounced position
angle sweep.}
\label{fig:cpsr2}
\end{figure}
\end{centering}

Fig \ref{fig:cpsr2} shows our most recent fully calibrated mean
profile, observed in 2004 May (MJD 53134) with the H-OH receiver and
CPSR2, at a centre frequency of 1341 MHz. There are a number of
striking differences when compared to Fig \ref{fig:cpsr1}, most
notably an overall broadening of the profile, which now has an
extended trailing component; loss of the leading ``bump'' and general
steepening of the position angle swing, which now has an identifiable
slope.

Pulsar profiles can be corrupted by systematic errors associated with
instrumentation. Typical effects include baseline artifacts due to
improper binning and radio frequency interference, insufficient
quantisation capabilities in the samplers (leading to the removal of
power near the pulse as the samplers attempt to maintain an optimum
mean level), and smearing due to insufficient time and frequency
resolution.  Fortunately, in the case of PSR J1141--6545 we have a
variety of instruments, each of which has sufficient time resolution
to over-sample the profile, and that give self-consistent results at
similar epochs. We would be concerned if observed changes only
coincided with equipment upgrades, but this has not been the case.

The mean pulse profile has changed significantly in the space of three
years. This is confirmed in the next two sections where we present an
analysis of data taken during (and before) the epochs presented in
Figs \ref{fig:cpsr1} \& \ref{fig:cpsr2}. We observe a smooth secular
change in the characteristics of the mean pulse.

\section{Evolution of the Total Intensity Profile}
\label{sec:tip_change}

To examine the evolution of PSR J1141--6545's mean pulse profile in
greater detail, we sum all the data within each observing session to
produce a set of 13 well spaced, high S/N profiles. These mean
profiles typically span one or two days, with total integrated times
of a few hours. Table 1 summarises the most important parameters of
each profile including observing system, start date, observing
frequency and S/N.

\begin{table}
\begin{center}
\begin{tabular}{c|c|c|c}
\hline
Instrument & Date (MJD) & Freq (MHz) & S/N \\
\hline
FPTM  & 51381 & 1318.25 & 125  \\
FPTM  & 51781 & 1413.50 & 791  \\
CPSR1 & 51922 & 1413.00 & 778  \\
FPTM  & 52087 & 1413.50 & 743  \\
CPSR2 & 52614 & 1405.00 & 1051 \\
CPSR2 & 52845 & 1341.00 & 427  \\
CPSR2 & 52902 & 1341.00 & 822  \\
CPSR2 & 52902 & 1341.00 & 1354 \\
CPSR2 & 52920 & 1341.00 & 390  \\
CPSR2 & 53109 & 1341.00 & 346  \\
CPSR2 & 53130 & 1341.00 & 728  \\
CPSR2 & 53133 & 1341.00 & 537  \\
CPSR2 & 53134 & 1341.00 & 553  \\
\hline
\end{tabular}
\caption{List of parameters associated with the 13 observations used
to characterise the secular evolution of the PSR J1141--6545 mean
pulse profile.}
\end{center}
\end{table}

\begin{centering}
\begin{figure}
\includegraphics[angle=270,scale=0.65]{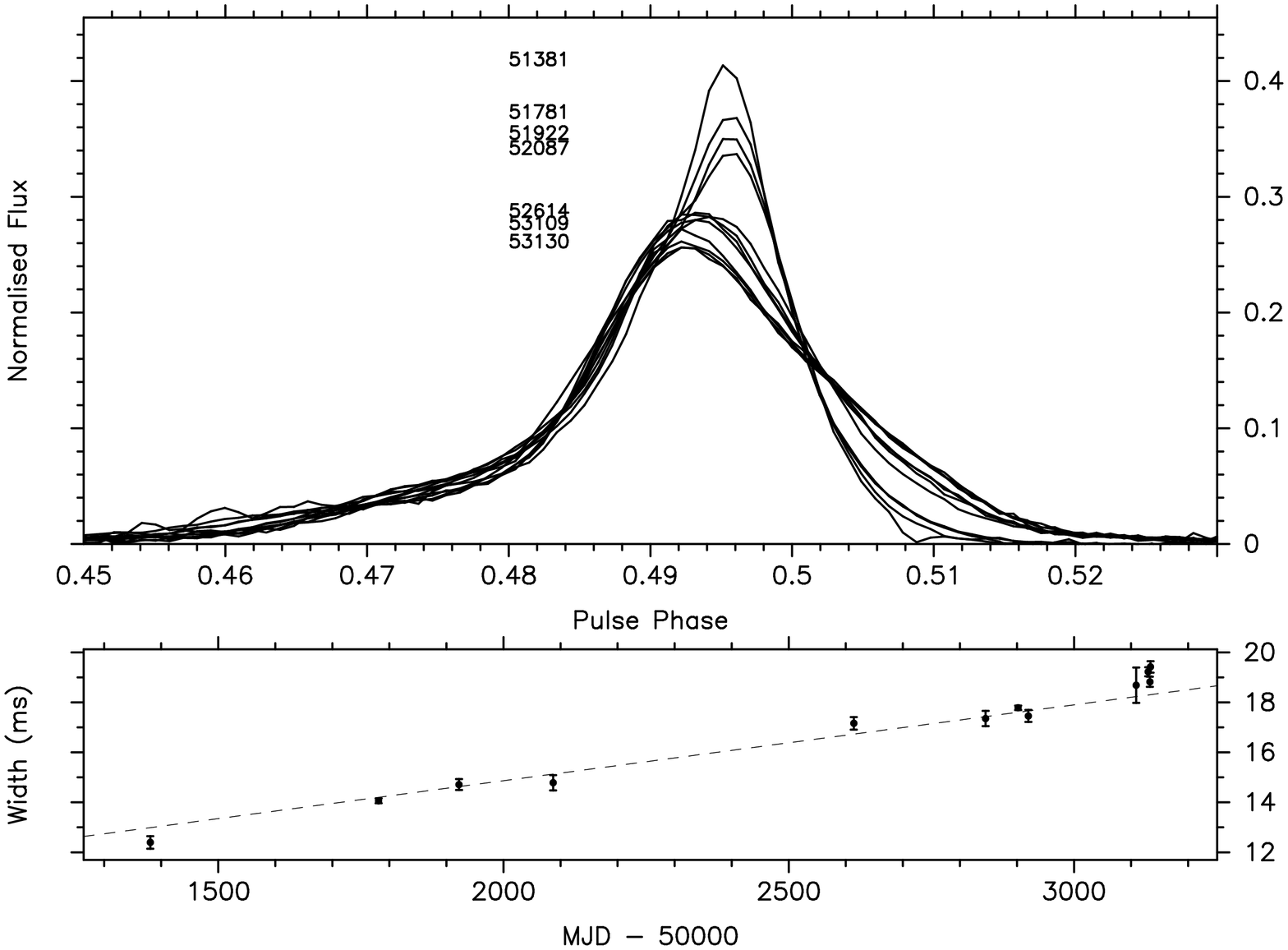}
\caption{The upper panel shows superposed mean total intensity
profiles from observations taken at various epochs during the past
five years. Each profile was observed at a wavelength near 20cm. The
MJD of each observation is marked on the plot, in line with the peak
of the corresponding profile. For example, at pulse phase 0.495, the
MJD of each observation increases monotonically with decreasing
amplitude. The flux under each profile in the plotted region has been
normalised to the flux under the earliest profile to allow direct
comparison, regardless of the amplitude scaling schemes used by
individual instruments. Each mean profile was aligned using an
ephemeris obtained from a global timing solution across the entire
data set, effectively maximising the cross-correlations between each
profile. The lower panel shows the evolution of 10 percent width (see
text) as a function of time, with one point for each profile in the
upper panel. The profile clearly broadens over the span of our
observations. Error bars are derived from consideration of the
root-mean-square (RMS) noise level in each profile and represent
1-$\sigma$ uncertainties. The line of best fit, obtained using a
linear least-squares method, is shown (dashed).}
\label{fig:mean}
\end{figure}
\end{centering}

\begin{centering}
\begin{figure}
\includegraphics[angle=270,scale=0.65]{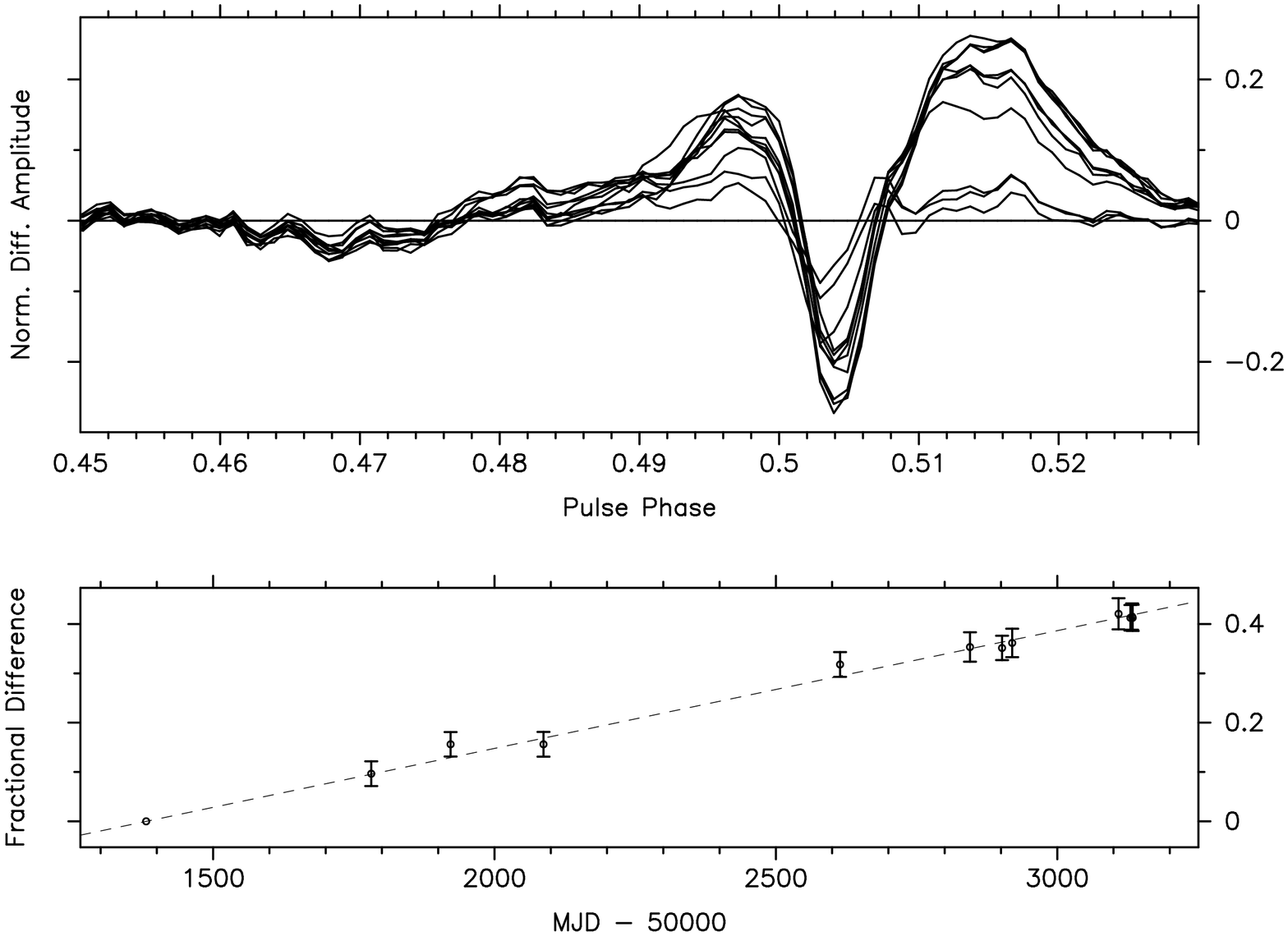}
\caption{The upper panel shows superposed difference profiles,
constructed from the data shown in Fig \ref{fig:mean}, using the
earliest profile as a standard template (shown as a horizontal line
through zero). The vertical scale is in units of the standard profile
flux, indicating that some individual components of the profile have
changed by up to 20 percent. There is a monotonic increase in
difference amplitude with time, at pulse phase 0.515 for example, the
amplitude increases with the MJD of the observation. The lower panel
shows the fractional difference between each difference profile and
the standard template, as a function of time. Each point represents a
single difference profile. The fractional difference is found by
summing the absolute values of the amplitude in each pulse phase bin
in the difference profile and dividing by the flux in the standard
template. Errors are based on measurements of the off-pulse
RMS. Whilst this technique is sensitive to changes in S/N as well as
morphology, we are confident that our choice of high S/N observations
and a narrow phase window allows the morphological information to
dominate. The line of best fit is shown (dashed), obtained using a
linear least-squares method. Note that the earliest profile has a
fractional difference (and error) of zero, providing a reference
point.}
\label{fig:diff}
\end{figure}
\end{centering}

We demonstrate that even though our points are not evenly spaced in
time, the data describe a clear trend in profile evolution (Figs
\ref{fig:mean} \& \ref{fig:diff}). The changes are so great that
visual inspection of the profiles can reveal much qualitative
information including an overall broadening, extension of the trailing
shoulder and smoothing of the leading edge (Fig \ref{fig:mean}). In
addition, we compute a quantitative measure of the width of each mean
profile using an algorithm that defines thresholds in pulse phase
based on where the flux under the pulse exceeds 10 percent of the peak
value for the first time, and drops below this value for the last
time. We choose the 10 percent threshold because it includes the
majority of the on-pulse region, thus incorporating the important
leading and trailing components of the pulse, which are seen to evolve
significantly. The secular trend remains if the level is set to 50
percent, however the $\chi^{2}$ of the linear fit worsens marginally
as we would expect from this narrower region of the profile.

A simple linear least-squares fit to the width data shows that the
rate of profile broadening is well approximated by a straight line
with slope 1.3 $\pm$ 0.06 ms yr$^{-1}$. Unlike PSR B1913+16, PSR
J1141--6545 has only a single pulse profile component. It is therefore
difficult to constrain the angular extent of the beam, or the angle
between the magnetic axis and the line of sight, using total intensity
information alone. A similar total intensity profile can be produced
by intersecting the centre of a narrow beam or the edge of a wider
beam, which introduces a degeneracy in the
interpretation. \citet{sta04} describe a method for determining the
full geometry of the system through the detection of a secondary
precession effect caused by orbitally modulated aberration.
Unfortunately we have been unable to detect this in PSR J1141--6545.
Given a sufficient time baseline, the profile variations should
eventually deviate from the currently observed linear trend. This will
also provide an opportunity to constrain the three dimensional
geometry of the pulsar system.

In order to characterise the rate of change in more detail, we perform
a difference profile analysis on our data set. This involves using
(arbitrarily) the first profile in the series as a standard template
whose amplitudes are subtracted from the remaining profiles after
their flux and alignment have been normalised to the standard.
Measurement of the remaining flux in the difference profile gives an
indication of how much the profiles vary across a particular epoch
(Fig \ref{fig:diff}). This method is similar to the technique of
principle component analysis performed on PSR B1534+12 by
\citet{sta04}, where only the mean profile and one orthogonal
component are taken into account. We use this numerical measure of
profile difference to examine quantitatively the rate at which
evolution is occurring.

Fig \ref{fig:diff} shows that the fractional difference trend is well
approximated by a straight line with slope $2.4 \pm 0.07
\times10^{-4}$ fractional difference units per day. The mean pulse
profile of PSR J1141--6545 is therefore changing at a steady rate of
approximately 9 percent per year. The profile evolution seen in Figs
\ref{fig:mean} \& \ref{fig:diff} is unlikely to be instrumental in
origin because it occurs smoothly over the entire time span, instead
of jumping discontinuously at the points when new hardware was
introduced. In addition, the instrumental upgrades always decreased
systematic smearing of the observed profile, whereas we observe the
profile width increasing with time.

\begin{centering}
\begin{figure}
\includegraphics[angle=270,scale=0.65]{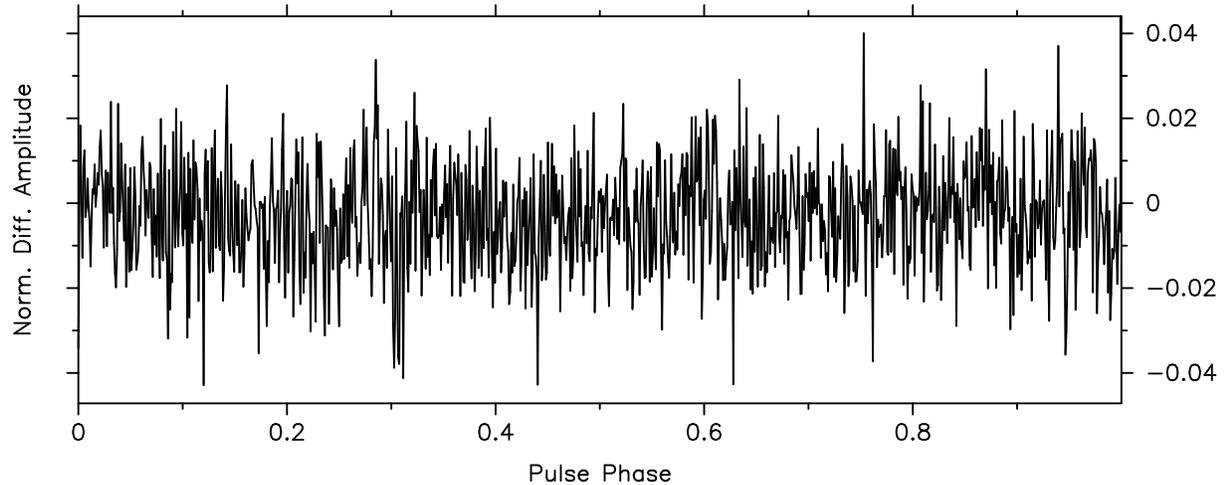}
\caption{Difference profile constructed from two CPSR2 observations
taken on MJD 53204 at two frequencies separated by 64 MHz. There is no
systematic morphological difference above the level of the noise that
might be expected if the pulsar profile was rapidly evolving as a
function of radio frequency.}
\label{fig:freqs}
\end{figure}
\end{centering}

Given that our observations span a frequency range of almost 100 MHz,
it is possible that intrinsic evolution of the profile shape with
frequency might contaminate the result. Such contamination is however
unlikely to be responsible for the observed secular trend in pulse
width, because as Table 1 shows, the changes in observing frequency
have not been linear in time. To demonstrate that PSR J1141--6545 does
not exhibit significant profile evolution over the range of observing
frequencies in our data, we compare two CPSR2 profiles observed on the
same day (MJD 53204) in July 2004. These two profiles were observed at
1405 MHz and 1341 MHz respectively and we analyse them in a similar
fashion to Fig \ref{fig:diff}, using the 1405 MHz profile as a
standard template and constructing a difference profile. Fig
\ref{fig:freqs} shows that the profile does not evolve significantly
across a bandwidth of 64 MHz, therefore we can be confident that
frequency evolution does not contaminate our result.

A number of different authors have reported observing the mean profile
of a pulsar changing in some way. Some slower pulsars have been
observed to emit in two or more ``modes'' of pulse shape, that each
remain stable for a time before switching rapidly to another mode
\citep{lyn71a}. In more recent years, various authors have reported
observing random profile variations in some of the millisecond
pulsars, most notably PSR B1821--24 \citep{bs97} and PSR J1022+1001
\citep{kxc+99}. None of the reported variations, or the mode changing
phenomenon, match the steady secular change we have observed in PSR
J1141--6545, which is itself a slow pulsar and therefore may not
suffer from the erratic variations seen in a small number of the
millisecond pulsars. It is still unclear whether or not random
variations in MSP profiles are intrinsic to the pulsar as a recent
analysis of PSR J1022+1001 observations made by \cite{hbo04} finds no
instabilities that might induce strange timing behaviour.

Can we hope to distinguish between geodetic precession and, say, free
precession of the pulsar? Pulse shape variations are attributed to
free precession in observations of PSR B1828--11 \citep{sls00} and PSR
B1642--03 \citep{slu01}. This evidence manifests in the form of
profile shape changes correlated to variations in the pulse arrival
times. These changes are cyclic, but with no clear {\it a priori} time
scale. The geodetic precession time scale is already well determined
to be near 265 years. From Fig 1. we would hope to see the emission
cone tilt at a rate $< 1.36$ deg yr$^{-1}$ and continue any secular
trend for decades unless we are at one of the two crossing points
encountered every geodetic precession period. The overwhelming
majority of slow pulsars exhibit no evidence for free precession
whatsoever. On the other hand, we expect any binary pulsar that has
received a misaligning kick to precess to some degree. We therefore
assert that geodetic precession is responsible for the observed
secular variation in the mean profile of PSR J1141--6545 and we
discuss this interpretation in section \ref{sec:discussion}. Long-term
monitoring of the precession will ultimately determine a time scale
and hence the true mechanism.

We now present an analysis of the polarimetry of this pulsar,
providing further evidence that our line of sight to the emission cone
is changing steadily with time.

\section{Evolution of Polarised Emission}
\label{sec:pol_change}

The standard interpretation of pulsar polarimetry is the rotating
vector model (RVM), put forward by \citet{rc69a}. This assumes a
dipolar magnetic field whose central axis is offset from the neutron
star rotation axis. As the emission cone sweeps past our line of
sight, the changing orientation between the observer and the magnetic
field is expected to produce a characteristic ``S'' shaped curve in
the measured position angle of any linearly polarised components. In
addition, the rate at which the linear polarisation vector rotates as
the beam crosses the observer is dependent on whether or not the line
of sight cuts close to the centre of the beam. Polarimetric
observations may therefore offer a sensitive indicator of both the
rate at which the beam is precessing past the observer and where in
the emission cone we are at the present time.

\begin{centering}
\begin{figure}
\includegraphics[angle=270,scale=0.7]{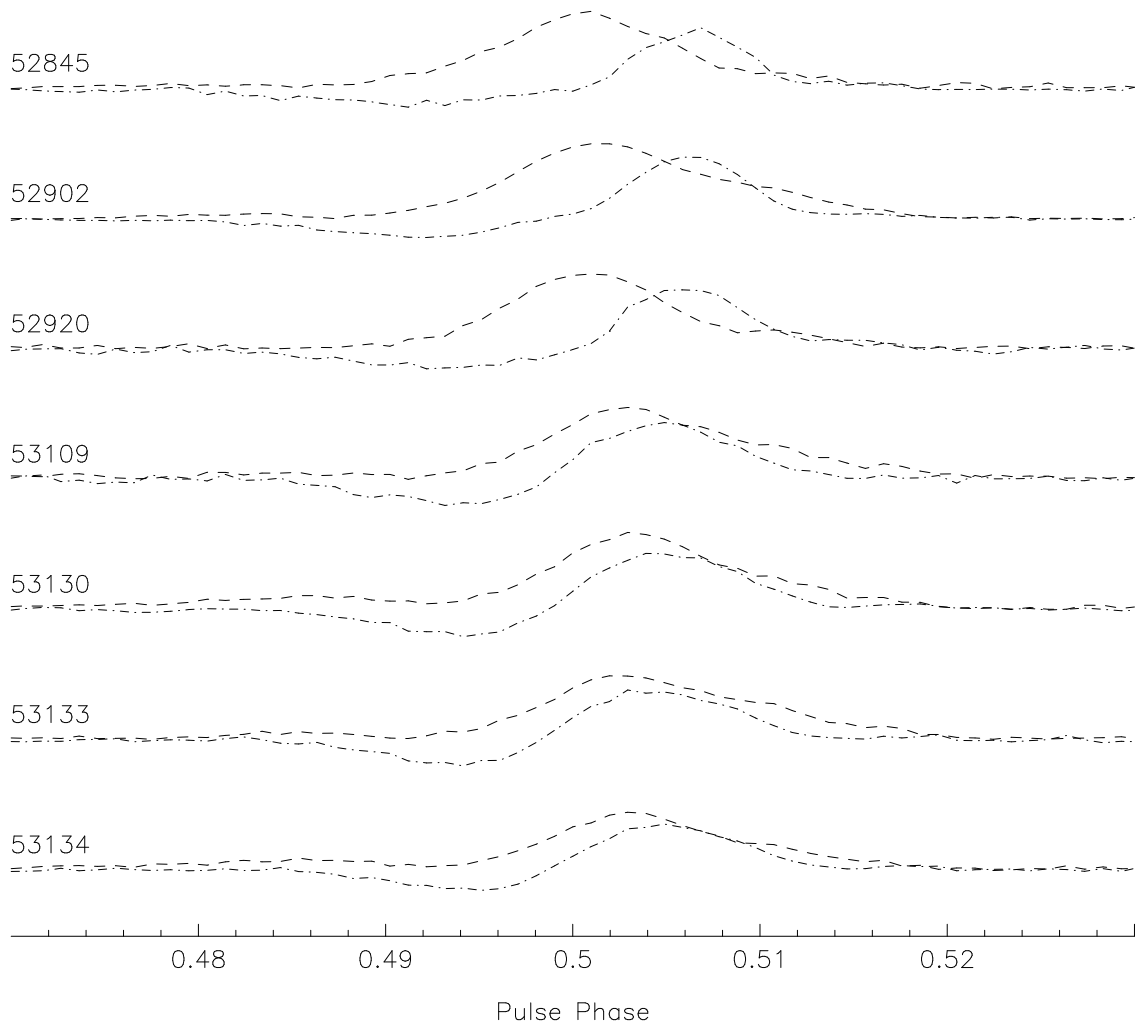}
\caption{Consecutive profiles of the polarised emission from PSR
J1141--6545, observed with CPSR2 over a period of 289 days beginning
in 2003 July (MJD 52845). The dashed line represents total linear
polarisation and the dashed-doted line represents total circular. The
vertical axis (unlabeled for simplicity) is in mJy, the vertical range
is kept constant across all sub-panels. All profiles have been
polarimetrically calibrated using a simple model of relative gain and
phase for two orthogonal linear receiver probes and flux calibrated
against Hydra A. The peaks of the linear and circular components can
be seen to move closer together as time progresses.}
\label{fig:pr}
\end{figure}
\end{centering}

First we consider the morphology of the polarised component of the
mean pulse. The polarimetric changes observed between Figs
\ref{fig:cpsr1} \& \ref{fig:cpsr2} are extreme. To convince ourselves
that rapid polarimetric evolution is taking place, we focus on the
most recent data with a high time density of observations and the same
pulsar back-end (CPSR2). The polarimetric capabilities of CPSR2 and
the PSRCHIVE scheme \citep{hvm04} have recently been verified by
comparison with published results with good agreement \citep{osh+04}.
Fig \ref{fig:pr} shows profiles of the polarised emission of PSR
J1141--6545, recorded using CPSR2 and ordered consecutively in time.

Fig \ref{fig:pr} shows that the polarimetric profile is certainly
changing, even over a period of less than 10 months. The linear and
circular emission appears to converge during this time. Significant
morphological differences in polarised emission appear between two
widely separated observations that also correspond to a change in
receiver, when the multibeam system was replaced by the refurbished
H-OH. Therefore it is still possible that the differences in polarised
emission could be instrumental in origin, although comparison with Fig
\ref{fig:cpsr1} suggests that position angle evolution is also
occurring. Observations over a longer time span are required to make
more conclusive (and perhaps quantitative) statements. Regular
polarimetric observations of this pulsar will be a high priority in
future years.

Next we attempt a more quantitative analysis of the position angle of
the linearly polarised component of the pulsar beam. In the formalism
of the RVM, measured position angle (PA, $\psi$), is presented as a
function of the angle between the spin and magnetic axes ($\alpha$),
the minimum angle between the magnetic axis and the line of sight
($\beta$) and pulse phase ($\phi$), as shown in Eq \ref{eq:rvm}.

\begin{equation}
\label{eq:rvm}
\tan(\psi(\phi) - \psi_{0})\, =\, \frac{\sin \alpha \sin(\phi -
\phi_0)}
{\cos\alpha\sin\delta - \sin\alpha\cos\delta\cos(\phi -
\phi_0)},
\end{equation} 

Here $\delta = \alpha + \beta$ is the angle between the spin axis and
the line of sight, as in section \ref{sec:1141}; $\phi_{0}$ is the
pulse phase of steepest PA swing and $\psi_{0}$ is a constant position
angle offset.

Unfortunately, application of this method to pulsars with narrow duty
cycles and shallow or complicated PA swings does not well constrain
$\alpha$ or $\beta$ independently. In common with the analysis
presented by \citet{klm+00} we find it impossible to fit the RVM model
to the early observations of PSR J1141--6545 with any degree of
confidence. More recent observations are better suited to a partial
RVM analysis, in that the recent PA behaviour includes a sweep in the
central region of the pulse profile. The narrow duty cycle still
restricts the applicability of the RVM and it is therefore impossible
to constrain $\alpha$. Under the assumption that the steepest PA
evolution is still providing an indication of the orientation of the
magnetic field lines with respect to both the spin axis and the line
of sight we have applied a simplification of the RVM model with the
sole purpose of determining the general evolution of $\beta$. Within
the RVM formalism the rate of change of PA as a function of pulse
phase has a maximum value, given in Eq \ref{eq:rvm2}:

\begin{equation}
\label{eq:rvm2}
\left(\frac{d\psi}{d\phi}\right)_{\rm max}\, = 
\, \frac{\sin\alpha}{\sin\beta}
\end{equation}

We have evaluated the gradient of the steepest PA swing in those
observations for which a straight line fit can be obtained from the
same central region of the pulse profile. This procedure could only be
applied to epochs between which the linear behaviour of the PA swing
was considerably wider in pulse phase than any possible translation,
allowing the profiles to be aligned by a suitable ephemeris. The
applicable observations are those obtained with CPSR2 since mid
2003. Earlier observations cannot be subject to this analysis as they
display PA behaviour which is too complicated. This fact alone
indicates that the detected emission represents a different cut
through the emission region than was evident in earlier observations.

\begin{centering}
\begin{figure}
\includegraphics[angle=270,scale=0.65]{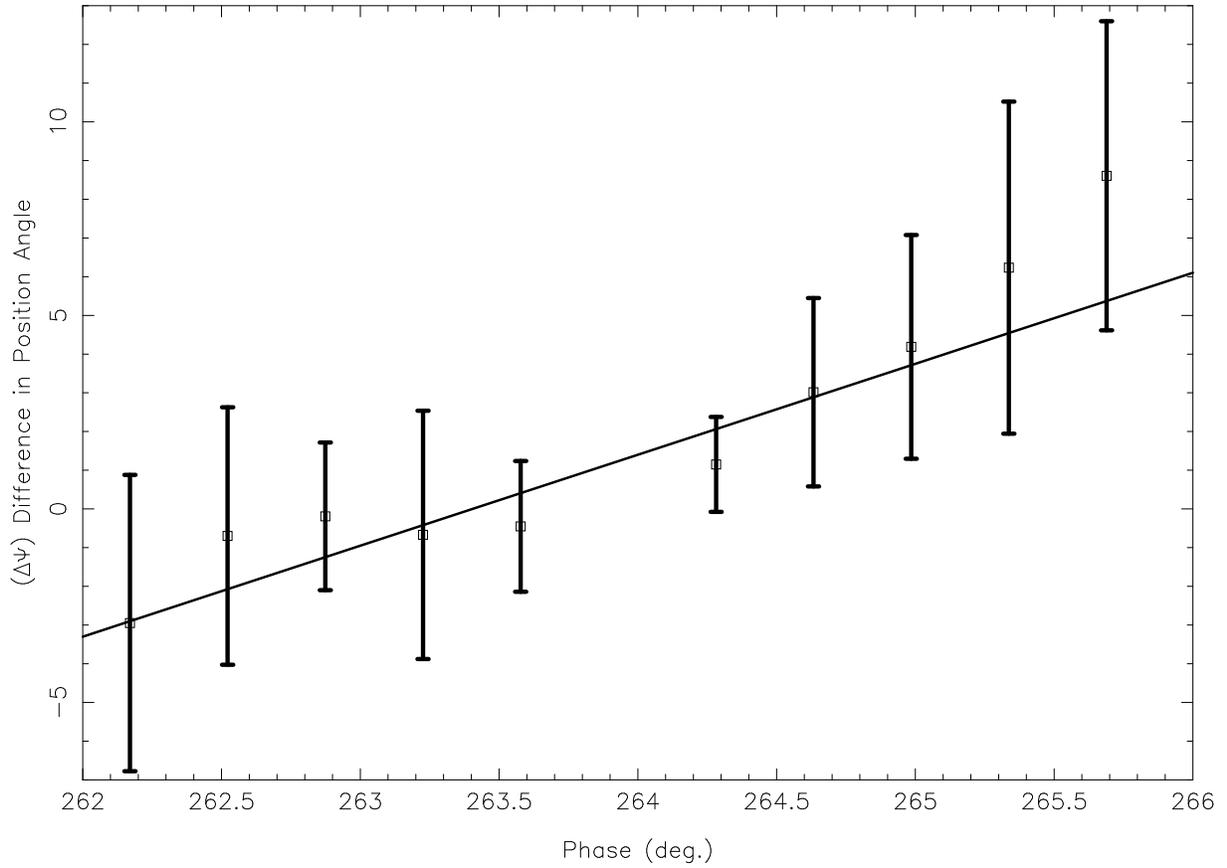}
\caption{The difference position angle (PA) as a function of
phase. The average PA from the mid-2004 epoch was subtracted from that
of the mid-2003 epoch. A straight line fit has been applied to the
residual. The steepening of the PA swing as a function of time is an
indication that the pulsar emission beam may be precessing into, and
not away from our line of sight.}
\label{fig:diffpa}
\end{figure}
\end{centering}

The data have been grouped into two epochs (2003 and 2004) separated
by approximately 0.7~years. Average position angle profiles were
formed from the constituent observations at each epoch. These average
PA profiles include observations from both 64 MHz observing bands and
within each epoch are separated by several days. Although there were
very slight variations between the PA profiles within each epoch,
there was no systematic trend. A linear least squares minimisation of
a straight line fit to the PA profiles demonstrates a significant
deviation in the gradient obtained at the two epochs. Epoch one (2003)
displays a gradient of --15.1 $\pm$ 0.3 degrees of PA swing per degree
of phase. Epoch two (2004) displays a gradient of --17.1 $\pm$ 0.3
degrees of PA swing per degree of phase. A straight line fit to a
difference PA profile, formed by the subtraction of the epoch two
profile from the epoch one profile, is presented in Fig
\ref{fig:diffpa}. The gradient of this difference fit is 2.3 $\pm$
0.4, which is consistent with the simple subtraction of the best fit
for epoch two from that of epoch one.

Eq \ref{eq:rvm2} was then evaluated for all $\alpha$ and a rate of
change of $\beta$ between the two epochs was determined. The
calculated value of $d\beta/dt$ is a strong function of $\alpha$ but
suggests that $\beta$ is increasing. The peak rate of change is 0.8
degrees yr $^{-1}$ (68 \% confidence), but this interpretation
requires the pulsar spin axis to be 90 deg from the magnetic axis,
which is unlikely if we assume the orientation is random. It is
therefore probable that $d\beta/dt$ is less than 0.8 degrees
yr$^{-1}$. The gradient of PA also indicates that $\beta$ is currently
negative for all possible $\alpha$. We therefore infer that the beam
is precessing into the line of sight at a rate less than 0.8 degrees
yr$^{-1}$. It should be noted that $d\beta/dt \equiv d\delta/dt$ as
the two angles are related by a constant offset. Thus we can use this
result, coupled with Fig \ref{fig:delta}, to make some general
statements about the unknowns in Eq \ref{eq:delta}. It is clear that
either the misalignment angle is smaller than approximately
30$^\circ$, or we are currently at a special precessional phase where
$d\delta/dt$ is changing rapidly and has not attained its maximum
value in the span of our observations.  Conversely, unless $\alpha$ is
near 0$^\circ$ or 180$^\circ$, we expect that the misalignment angle
is greater than approximately 15$^\circ$.

The mean flux of the pulsar in these observations is 4.0 $\pm$ 0.5
mJy. There does not appear to be a significant trend in mean flux
within the past year, but this may change as the time baseline of flux
calibrated data grows.

\section{Discussion \& Conclusion}
\label{sec:discussion}

Our observations indicate that the line of sight to the emission cone
of PSR J1141--6545 is precessing deeper into the core, producing a
steeper position angle swing. Depending on the chosen beam model, this
could also explain why the pulse profile is seen to increase in
width. Our result has implications both for the study of pulsar
emission beams and for the detection rates of relativistic pulsars in
large-scale surveys. PSR J1141--6545 is the slowest pulsar (by an
order of magnitude) for which geodetic precession is observable,
providing a unique means of examining the emission cone of a normal
(un-recycled) pulsar. Further polarimetric observations, extended time
baselines or the detection of orbitally modulated aberration may allow
determination of our present location in the emission cone and the
geometry of the beam as a whole. Given the expected precession period
of order 265 years and the fact that we seem to still be moving
towards the central axis of the beam, it is possible that this pulsar
may only have precessed into view within the past few decades. This
might explain its non-detection in early pulsar surveys with flux
limits well below the required threshold \citep{jlm+92, mld+96}. If
this is the case, it could be argued that surveys of the sky for
relativistic pulsars should continue on a regular basis. By their very
nature, the most interesting objects are likely to be visible for the
least amount of time.

%% Included in this acknowledgments section are examples of the
%% AASTeX hypertext markup commands. Use \url without the optional [HREF]
%% argument when you want to print the url directly in the text. Otherwise,
%% use either \url or \anchor, with the HREF as the first argument and the
%% text to be printed in the second.

\acknowledgments

The Parkes radio telescope is operated by the Australia Telescope
National Facility on behalf of the Commonwealth Scientific and
Industrial Research Organisation (CSIRO). We are grateful to Ruth
Musgrave for assistance with data reduction and Willem van Straten and
Haydon Knight for assistance with observations and software
development. AWH is the recipient of an Australian Post-graduate Award
(APA) and CSIRO top-up allowance and thanks Claire Trenham for support
and encouragement.

\bibliographystyle{apj}
\bibliography{journals,psrrefs,modrefs,awh}

\end{document}